\documentclass[12pt,preprint,epsfig]{aastex}
\usepackage{emulateapj5}

\slugcomment{}

\shorttitle{}
\shortauthors{D'Onghia $\&$ Burkert}

\begin{document}


\title{Bulgeless Galaxies and their Angular Momentum Problem}


\author{Elena D'Onghia\altaffilmark{1,2}\\}
%\affil{Max-Planck-Institut f\"ur extraterrestrische Physik,  85748
%  Garching, Germany}
%\email{donghia@usm.uni-muenchen.de}

\author{Andreas Burkert\altaffilmark{2}\\}
%\affil{University Observatory Munich,
%  Scheinerstrasse 1 81679 Munich, Germany}
%\email{andi@usm.uni-muenchen.de}

%\and
\altaffiltext{1}{Max-Planck-Institut f\"ur extraterrestrische Physik,  85748
  Garching, Germany, donghia@usm.uni-muenchen.de}
\altaffiltext{2}{University Observatory Munich,
  Scheinerstrasse 1 81679 Munich, Germany, andi@usm.uni-muenchen.de}


%\altaffiltext{1}{Visiting Astronomer, Cerro Tololo Inter-American Observatory.
%CTIO is operated by AURA, Inc.\ under contract to the National Science
%\altaffiltext{2}{Society of Fellows, Harvard University.}
%\altaffiltext{3}{present address: Center for Astrophysics,
%    60 Garden Street, Cambridge, MA 02138}
%\altaffiltext{4}{Visiting Programmer, Space Telescope Science Institute}
%\altaffiltext{5}{Patron, Alonso's Bar and Grill}


\begin{abstract}

The specific angular momentum of Cold Dark Matter (CDM) halos in a 
$\Lambda$CDM universe is investigated. Their dimensionless specific angular momentum
$\lambda'=\frac{j}{\sqrt{2}V_{vir} R{vir}}$ with $V_{vir}$ and $R_{vir}$ the 
virial velocity and virial radius, respectively depends strongly on their merging histories.
We investigate a set of $\Lambda$CDM simulations and explore the specific
angular momentum content of halos formed through various merging histories. 
Halos with a quiet
merging history, dominated by minor mergers and accretion until the present epoch,
acquire by tidal torques on average only 2\% to 3\% of the 
angular momentum required for their rotational support ($\lambda'=0.02$). This is in conflict
with observational data for a sample of late-type bulgeless galaxies 
which indicates that those galaxies reside in dark halos with exceptionally high values of
$\lambda' \approx 0.06-0.07$. Minor mergers and accretion preserve or slowly
increase the specific angular momentum of
dark halos with time. This mechanism is however not efficient enough in order to explain
the observed spin values for late-type dwarf galaxies.
Energetic feedback processes have been invoked to solve the problem that gas
loses a large fraction of its specific angular momentum during infall.
Under the  assumption that  dark halos hosting bulgeless galaxies
acquire their  mass  via  quiescent accretion,
our results indicate yet another serious problem: the specific angular momentum gained
during the formation of these objects is not large enough to explain their observed 
rotational properties,
even if no angular momentum would be lost during gas infall.
%Our results indicate that cosmological  models of bulgeless galaxy formation
%have an even more severe problem as even without any angular momentum loss the
%specific angular momentum gained through smooth merging and accretion will be a factor of 3 smaller
%than observed.

\end{abstract}


\section{Introduction}
The dynamical structure of disc galaxies is dominated by angular momentum.  Therefore,
understanding the origin of angular momentum in these systems
is crucial in any theory of galaxy formation.
In the current paradigm for structure formation, dark matter is
assumed to be cold and collisionless and luminous galaxies form by
gas infall into  dark matter halos,  which grow by gravitational accretion 
and merging in a hierarchical fashion (White $\&$ Reese 1978).
Fall $\&$ Efstathiou (1980, hereafter FE80) proposed that
the sizes of galactic disks are linked to the angular momentum of their
parent dark matter halos. This theory is able to produce disks with 
sizes that are in agreement with observations, if  
the gas initially had the same specific angular momentum 
as dark matter halos show today and if the gas preserved its
specific angular momentum during the protogalactic collapse phase.
The angular momentum of galaxies results from torques due to tidal interactions
with neighbouring structures, acquired early, before the halo decoupled from the Hubble
expansion.

Many models for the formation of galactic disks have been proposed, based on the picture
of FE80. Most of them incorporate the mass accretion
history of halos and are able to reproduce many properties of observed disk galaxies
(e.g. Mo, Mao $\&$ White 1998; Firmani $\&$ Avila-Reese 2000; van den Bosch
2000). However, in  these models, the angular momentum
of the dark matter halos is assigned without accounting for their merging 
history. Recent results from numerical N-body simulations
have pointed out that the effect of major mergers is
to increase the mean angular momentum content of the halos (Gardner
2001, G01 hereafter; Vitvitska et al. 2001). This is explained by 
the orbital angular momentum of the merging halos which dominates
the final net angular momentum of the remnant (G01).
However, numerical simulations that incorporate gas dynamics
have difficulties to make realistic disk galaxies in the current
cosmological paradigm. In most simulations,
the disks are smaller, denser and have much lower angular
momenta than observed disk systems.  Simulations show that galaxies
are built up by merging of baryonic subclumps, rather than smooth 
accretion of gas. Most of the gas cools at the centre of
subhalos and spirals toward the center of the parent halo, transferring
orbital angular momentum to the surrounding dark matter (e.g.  Navarro $\&$ 
Steinmetz 2000). 
More acceptable fits to real disk systems can be found if heuristic
prescriptions of stellar feedback are included in the 
simulations (Sommer-Larsen et al. 2003; Abadi
et al. 2003, Robertson et al. 2004). However, even in these simulations the disk systems 
typically contain denser and more massive bulges than observed in late-type
galaxies. 
%making clear that there still exists a specific angular momentum problem.

Most of the previous work has focused on angular momentum properties of
halos that had at least one dominant major merger during their evolution. 
%However, major merger events tend to destroy disks, producing  
%spheroidal stellar systems, like bulges or early-type galaxies (see e.g.  review by
%Burkert \& Naab 2003, BN03 hereafter). 
%Not much work has been devoted to explore the angular momentum
%properties of halos that host pure disk galaxies, or bulgeless
%galaxies and that never experienced a major merger.
In this Letter we explore the angular momentum properties 
of halos that did not experience any major merger since
redshift 3 and that are in principle good candidates to
host bulgeless galaxies.
We demonstrate that even these galaxies have an angular momentum problem that is
directly related with the spin of their dark halos and cannot be solved easily by energetic
feedback processes.

\section{Simulations}

We performed three simulations within a $\Lambda$CDM cosmological universe
with $\Omega_0=0.3$, $\Omega_{\Lambda}$=0.7, $h$=H$_0/70$ 
km s$^{-1}$ Mpc$^{-1}$ and  $\sigma_8=0.9$.
The simulated volume was 15 $h^{-1}$ Mpc box size in all runs.
Each simulation was performed using the publicly available version of the smoothed
particle hydrodynamics (SPH) code GADGET (Springel, Yoshida $\&$ White
2001). All runs started at redshifts
sufficiently high to ensure that the absolute maximum density contrast 
$|\delta|\leq 1$. The simulations began from a spatially uniform grid of 
$128^3$ equal-mass particles with Plummer--equivalent softening comoving 
length of $10 h^{-1}$ kpc.  The particle masses was $1.34 \times 10^8$ $h^{-1}$ M$_{\odot}$. 

We identify halos with the classic friend-of-friend (FOF) method
using a linking length that corresponds to the mean interparticle 
separation at that density contour that defines the virial radius of an isothermal sphere,
$b=[n\rho_{vir}(z)/<\rho(z)>/3.0]^{-1/3}$=0.15, where $n$ is the particle number density
and $\rho_{vir}(z)$ is the corresponding virial density. 
All halos at z=0 with masses between M$=10^{11}-10^{12}$ $h^{-1}$ M$_{\odot}$,
containing at least 1000 particles are included in the analyses.
This restriction limits the influence of numerical effects 
on global halo properties. Disks with non-negligible bulge components of
$B/(B+D) \geq 0.1$, with B and D the bulge and disk mass, respectively, require
a major merger which involved at least 10\% of their final mass  (i.e. at least 100 particles).
Our focus on halos with final particle numbers of at least 1000 therefore guarantees
that we don't miss any relevant major merger events due to softening. 
We also did not find a dependence of our
results on the adopted mass range of dark halos at $z=0$. 
Therefore, neglecting 
smaller halos will not affect our conclusions.

\section{Bulge to Disk ratio vs. Spin parameter}

The angular momentum of a galaxy, $J$, is commonly expressed in terms of the dimensionless
spin parameter $\lambda=J\sqrt{|E|}/$GM$^{5/3}$where $E$ is the total energy, 
and M is the total mass.
%The value of the spin parameter represents the ratio between the actual
%angular velocity of the system, divided by that
%needed to support the system purely by rotation.
The halo spin parameter distribution in simulations is found to be well approximated by the 
log-normal function:\\
\begin{equation}
p(\lambda)\rm{d}\lambda=\frac{1}{\sigma_{\lambda}\sqrt{2\pi}}
\rm{exp}\Big[-\frac{\rm{ln}^2(\lambda/<\lambda>)}{2\sigma^2_{\lambda}}\Big] \frac{\rm{d}\lambda}{\lambda}
\end{equation}  
In practise, it is more convenient to use the modified spin parameter
(Bullock et al. 2001): $\lambda'=j/(\sqrt{2}V_{vir}R_{vir})$
%\begin{equation}
%\lambda'=\frac{j}{\sqrt{2}V_{vir}R_{vir}} \\
%\end{equation}
where $j=J/M$ is the specific angular momentum.  
For a $\Lambda$CDM cosmology the log-normal parameters are found to 
be $<\lambda'>=0.035\pm 0.006$
and $\sigma_{\lambda'}$=0.5-0.6 (Bullock et al. 2001).

We trace each identified halo backward in time, following the mass
of the most massive progenitor as a function of redshift during $0<z<3$. 
Our results don't change if we include major mergers at higher redshifts since
they in general don't include a last mass fraction. In addition,
the gas content of halos with such a high redshift is probably
large and the result of a major merger is very unclear (Bullock et al. 2001).
We start when the mass of the most massive progenitor
is 30\% of the z=0 mass. Assuming instead 50\% of the halo mass at present time does 
not change the results.  To identify mergers, 
we denote a halo as a major merger remnant if at some time 
during $0<z<3$ its major progenitor
was classified as a single group in one output, but two separate groups with a mass ratio$\leq 4:1$
in the preceeding output. 
The critical mass ratio of 4:1 for a major merger event
is based on results of N-body simulations which show
that at least such a mass ratio is required in order to produce a
spheroidal elliptical galaxy with a surface density profile that agrees
with observations of early-type galaxies (Burkert \& Naab 2003, BN03 hereafter). 
We find that 1:1 and 2:1 mergers are much more
frequent than 3:1 or 4:1 mergers, in agreement with earlier results by Khochfar
$\&$ Burkert (2004), based on an analyses of a much larger cube.  
As discussed earlier, we expecte our identified halo population to be complete at z=0
for masses  $\geq 10^{11}$ M$_{\odot}$. A slight incompleteness for 4:1 mergers
right at this mass limit is expected which should however not affect our results,
as 4:1 mergers in this limited mass range are very rare.
For each halo we identify the time of the last major merger.
We assume that accretion events and minor mergers encrease the mass of the
disk component whereas major mergers convert all the available
gas into stars and destroy disks, forming a stellar spheroid (BN03). 
Neglecting galactic
winds and assuming a universal baryon fraction for all infalling substructures, the
final ratio between bulge mass B and the sum of bulge and disk mass (B+D) is
given by the ratio of the dark halo mass at the time of the last major merger,
divided by the dark halo mass at $z=0$.

Fig.1 shows the B/(B+D) ratio at z=0 of halos analysed in one of  our
simulations as a function of their spin parameter $\lambda'$.
For better visualisation we just plot the results of
one run as the other two runs show similar trends. 
Different symbols are used to denote halos that have their last major
merger at different epochs (see legend in Fig.1). The plot shows
that the halo spin parameter is a function of the ratio between bulge mass
and total baryonic mass. Many halos with 
high B/(B+D) ratios also show higher $\lambda'$ values than the average 
$<\lambda'> \approx 0.035$ in the log-normal distribution
(filled symbols in Fig.1). Most of these halos experienced a recent
major merger in the redshift range $0<z<1$.
Halos that have not experienced any major mergers from 0$<z<3$, 
show a log-normal 
distribution with $<\lambda'> \approx 0.023$, significantly lower than 
the value of 0.035 corresponding to 
the average referred to all halos. This finding confirms previous results
that the 
last major merger event is affecting the final halo spin parameter 
distribution (G01; Vitvitska et al. 2002, Peirani et al. 2002).
In order to test how reasonable is to set to zero in Fig.1 
the bulge masses of halos that have not experienced any major merger from 0$<z<3$, 
we traced back the ``quiescient'' halos, attributing bulge masses different from zero to
those halos with the last major  merger occurs before z=3.
They show bulge mass B between 1\%-7\% of the sum of bulge and 
disk mass (B+D).
Since we are interested in bulgeless galaxies,
we focus here on halos with a quiet
merging history dominated by minor mergers and accretion which corresponds to
roughly 10\% of all halos. 
%Properties of the
%other halos will be explored in a forthcoming paper.
It would be interesting to compare the theoretically predicted frequency
of bulgeless galaxies with observations.  This would however require a much larger
survey with higher resolution and larger box sizes which we postpone to a subsequent paper.

In the general picture of FE80,
disks should have the same distribution of total specific
angular momentum as the dark matter halos, with 
%Hence they should have 
the same value for the spin parameter, $\lambda'_{disk} \approx \lambda'$. 
This is expected because all the
material experiences the same external torques during the early expansion phase
before separating into two distinct components as a result of dissipative
processes during the collapse phase.
Numerical simulations have shown that gas and dark matter have identical 
angular momentum distributions after the protogalactic collapse if cooling
is ignored (van den Bosch et al. 2002). 
%of disk galaxy formation
%show that this assumption seems to be violated. 
During the protogalactic collapse phase, if cooling is included, baryons lose most of their initial 
angular momentum to the dark matter by
dynamical friction (Navarro $\&$ Steimetz 2000), resulting in galactic disks
that are smaller than observed.  Suppression of early cooling of the gas
by energetic heating e.g. through supernovae is invoked as a mechanism to prevent this 
angular momentum catastrophe.
We find however that disk-dominated late-type galaxies inhabiting
halos that have not experienced a major merger have a
distribution of $\lambda'_{disk}$ that peaks around a
value of 0.023  which is substantially smaller than expected from observed 
rotation curves.  To demonstrate that we use the results of
van den Bosch, Burkert $\&$ Swaters (2001, BBS hereafter) who determined spin
parameters for galactic disks of 14 late-type bulgeless dwarf galaxies. 
Fig.1 shows the spin parameters of their galaxies, 
assuming a mass-to-light ratio of unity in the R band. 
The result is surprising: none of our models lies in the 
region of the diagram covered by the observational data.
   
Fig.2 (top panel)  shows the probability distribution of the spin
parameter of halos that did not experience any major mergers since z=3
(dashed region) and compares it with the normalized probability distribution
of $\lambda'_{disk}$ for the sample of galaxies measured by BBS. 
The dwarf galaxies show a distribution that follows a log-normal with 
an average value for $\lambda'_{disk} \approx 0.067$, and a dispersion of
$\sigma_{\lambda'}\approx 0.31$. This is a factor of 3 larger than predicted
by the numerical simulations. Again, galactic disks would be too small. 
%However,
%this time, the problem cannot be solved by feedback processes which suppress
%the loss of specific angular momentum of clumpy infalling gas.
It is remarkable that multiplying the $\lambda'$ values of simulated
halos by a factor 3.15 (dotted histogram in Fig.2) reproduces the peak
and dispersion of the observed $\lambda'_{disk}$ distribution quite well.
Thus, the discrepancy could be due to a systematic
underestimation by a factor of 3 of the product R$_{vir} \cdot$V$_{vir}$  in the BBS  sample.
However, if R$_{vir} \cdot$V$_{vir}$ is increased of a factor of 3, then M$_{vir}$
increases of a factor of $3^{1.5} \approx 5.2$, which implies a factor
5 smaller baryon fraction. The inferred baryon fraction
in the BBS sample is already quite small (see Fig.3 of BBS). This solution therefore appears
to be very unlikely.  On the other hand, the dispersion $\sigma_{\lambda'}$ derived from the
observations is in good agreement with the dispersion 0.36, derived for
disk galaxies, based on an analysis of 1000
Sb-Sdm galaxies (de Jong $\&$ Lacey 2000). 
In the bottom panel of Fig.2 we show the probability distribution 
of the spin parameter of the entire halo population (thin solid line).
The distribution is a log-normal with $\lambda_{0}'=0.034 \pm 0.005$ and 
a width $\sigma=0.45 \pm 0.03$. The value for $\sigma$ is somewhat smaller than 
previous analysis (G01; Bullock et al. 2001). It is however still within the statistical
uncertainties that will in future work be reduced by analysing larger cubes.
The figure also shows the frequency of  Sub-populations are also included  in the  total halo  
histogram showing the result if major merger are defined as mergers with mass ratios
$>1:4$ (thick solid line), $>1:3$
(dotted line) and $>1:2$ (long-dashed line), respectively. Note that the spin distributions
of the late-type dwarfs in the bottom panel of Fig.2 is in disagreement
with all theoretical distributions.
Another point of concern could be a dependence of rotational properties on galaxy mass. 
In our simulations all analysed halos have masses
between M$=10^{11}-10^{12}$ $h^{-1}$ M$_{\odot}$ which is a factor of 10 to 100 larger
than the inferred virial mass of the observed dwarf galaxies of the BBS sample.
One might argue that halos of lower mass have higher spin parameters,
although it is well known that the spin parameter
distribution of halos is independent of mass (Lemson $\&$ Kauffmann 1999).
We have investigated this question and find no correlation between $\lambda'$
and $M_{vir}$ of the quiet halos in our simulations.

\section{Discussion and Conclusion}

If bulgeless galaxies have not experienced
any major mergers during their evolution our analysis shows that
their dark halos  are characterized by systematically lower 
angular momentum than observed. Halos without major mergers 
acquire their specific angular momentum through tidal torques 
in the early epochs of evolution (Barnes $\&$ Efsthatiou 1987),
when the density contrasts were small, in
accordance with the prediction of the linear theory. 
In Fig.3, top panel, the spin parameter evolution of the
major progenitors of two representative halos is shown. Due to  the lack of any 
major merger event, there is no sharp increase in $\lambda'$ which is
characteristic for major mergers. Instead,
$\lambda'$ is almost constant and slightly increasing with
time, proving that minor mergers and accretion does not affect
or significantly  decrease $\lambda'$, in contradiction with claims of 
Vitvitska et al. 2002. The bottom panel shows that
the halos aggregate mass gradually by  accretion of small subhalos. 
%We have resimulated one of these halos with higher resolution, 
%and verified that its spin evolution is independent of resolution.
Robertson et al. (2004) have
recently    presented  the  results  of  a   cosmological hydrodynamic
simulation where they form an extended, bulge-less disk galaxy.
In our sample of "quiescent" halos, there is one case with high spin value ($\lambda'=0.05$ at z=0), increasing
more than a factor of 2 from z=0 and z=3 and no major merger during that time. In its accretion history, 
mergers with mass ratio 1:5, 1:6 seem to be enough to spin up the halo, although they are 
unlikely in CDM scenario. Thus, of course this object could be 
a good candidate to be a bulgeless galaxies and we are resimulating it with 
higher resolution (D'Onghia et al. in preparation), but cannot
be considered as a representative ``quiescient'' halo. Robertson et al. 2004
could have simulated an object with similar properties, although the authors don't report the spin
parameter value or the accretion history  of their object.

The net result seems to be that  tidal torques generate
halos with typical spin parameters of $\lambda' = 0.02$
whereas data for bulgeless galaxies indicate halos with values of $\lambda' = 0.06 - 0.07$.
One might worry that our result is affected by the adopted small cosmological 
volume of 15 $h^{-1}$ Mpc box size, that might suppress torques by 
by larger-scale structures. 
However tidal forces scale as $F \propto 1/r^3$. Compared to the forces that are
generated by structures at a Mpc distance,
structures located at distances of 10 Mpc need to have the mass $10^3$ larger than
at 1 Mpc to have the same effect on protogalaxies. At 100 Mpc, the mass of the 
structure would have to be even$10^6$ times larger to torque protogalaxies efficiently,
which is unlikely.
In addition, our results are in agreement with the models of G01 
for a larger volume of 100 Mpc box size.

At the moment it is not at all clear how one can reconcile the observations 
and theory.
It is known that the spin parameter distribution for the
collapsed objetcs is insensitive to the shape of the initial power
spectrum of density fluctuations, to the environment
and the adopted cosmological model (Lemson $\&$ Kauffmann 1999). 
A modification of the nature of dark matter
does not seem to solve the problem either. Recent works show that warm dark
matter  halos have
systematically smaller spins than their counterparts in
$\Lambda$CDM (Knebe et al. 2002; Bullock, Kravtsov $\&$ Colin 2002),
although in this scenario the presence of pancakes could provide more
efficient torques on the protohalos. 
%If dark matter is assumed to be collisional, there exists still no
%reason why this properties should increase the specific angular momentum
%of the baryonic component.
Feedback was invoked as a mechanism to prevent
the process of drastic angular momentum loss of infalling gas.
(BBS; Maller, Dekel $\&$ Somerville 2002).
However, we have shown here that the dark
halos that experienced no major mergers have already too low an angular momentum to
produce the observed disks and no feedback process is know that would increase the
the specific angular momentum of the gas. The origin of extended bulgeless disk galaxies
remains a puzzle.

\acknowledgments
We are grateful to the referee for his suggestions, to 
G.Murante for his help with the merging 
tree, T.Abel, F.van den Bosch, A.Dekel, B.Moore, J.
Primack, S.White for stimulating discussions and comments.

\clearpage


\begin{figure}
\plotone{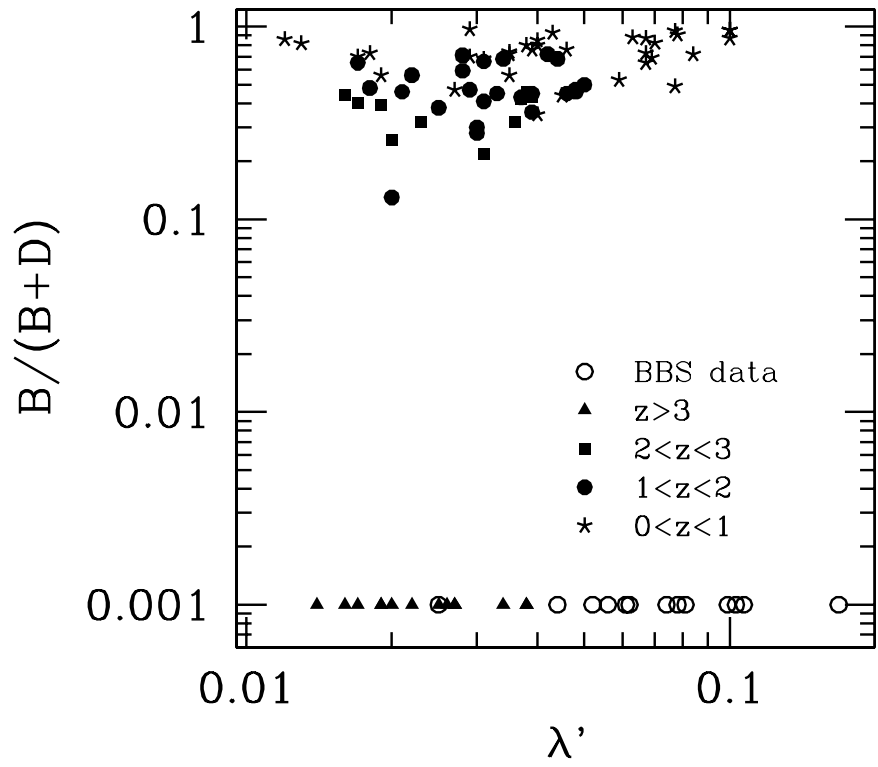}
\caption{The bulge mass (B) compared to the total baryonic mass (B+D) 
is plotted for each halo against its spin parameter computed 
at z=0. Different symbols mark halos that had the last major merger at different epochs. 
The spin parameter values measured for 14 late-type dwarf galaxies
by van den Bosch, Burkert $\&$ Swaters (2001)(BBS) are also plotted.} 
%\label{fig1}}
\end{figure}

\clearpage 
\begin{figure}
\plotone{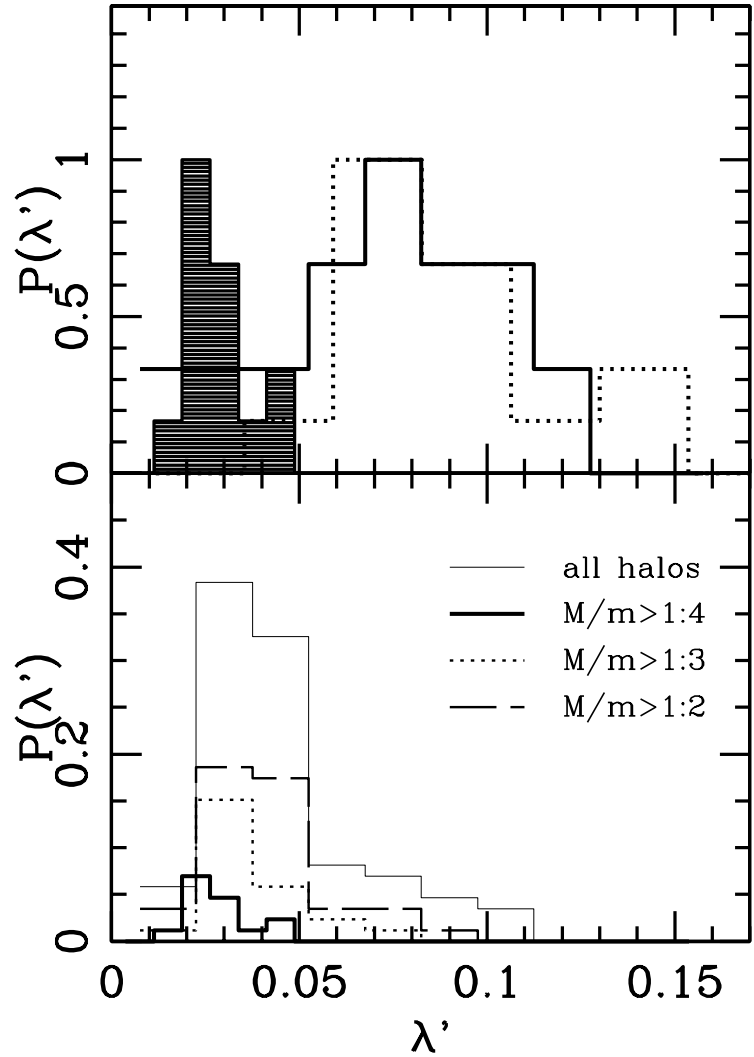}
\caption{{\it Top Panel}. The probability distribution of the spin parameter of halos
that did not experience any major merger since z=3 (dashed region)
is compared to the distribution inferred from the BBS data (solid line). The dotted line shows the result
if the $\lambda'$-values of the simulated halo distribution would be multiplied by a 
factor of 3.15.  {\it Bottom Panel}.  The probability distribution 
of the spin parameter of the entire halo population (thin solid line).
Sub-populations are also plotted if one assumes that the definition of a major mergers requires
mass ratios of $>1:4$, $>1:3$, and  $>1:2$, respectively, since
z=3.}
\end{figure}

\clearpage 
\begin{figure}
\plotone{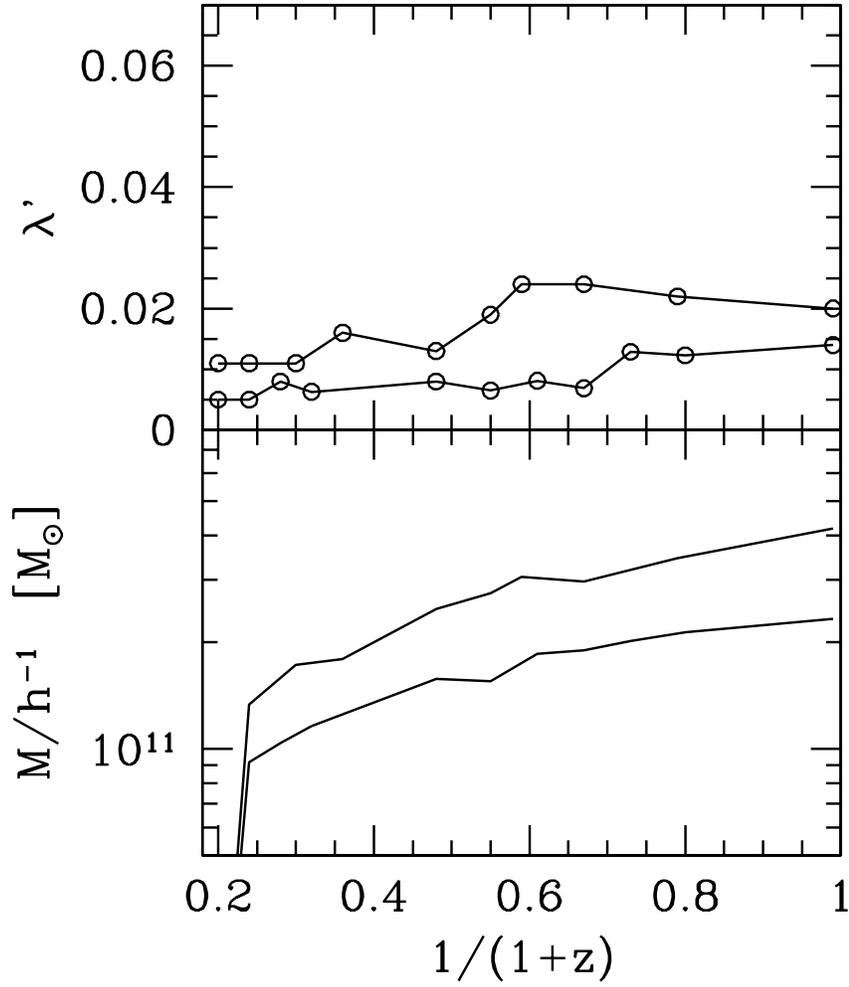}
\caption{The spin parameter evolution of the major progenitors
of two representative halos that did not experience any major merging since z=3 (top panel)
and their corresponding mass accretion history (bottom panel) is shown.}
\end{figure}

%\begin{figure}
%\plottwo{f2a.eps}{f2b.eps}
%\caption{This is an example of a multipart figure with a long figure caption 
%that must be set as a paragraph.  The processor has to buffer the text of the
%caption, so it is good not to be too wordy, but that would make for
%poor communication as well.\label{fig2}}
%\end{figure}

\clearpage

\end{document}